# AN ANALYSIS OF DIALOGUE REPAIR IN VIRTUAL VOICE ASSISTANTS

Matthew C. Galbraith, Mireia Gómez i Martínez

Department of Applied and Theoretical Linguistics, Universitat Pompeu Fabra

**Abstract**

Language speakers often use what are known as repair initiators to mend fundamental disconnects that occur between them during verbal communication. Previous research in this field has mainly focused on the 'human to human' use of repair initiator. We proposed an examination of dialogue repair structure wherein the dialogue initiator is human and the party that initiates or responds to the repair is a virtual assistant. This study examined the use of repair initiators in both English and Spanish with two popular assistants, Google Assistant and Apple's Siri. Our aim was to codify the differences (if any) in responses by voice assistants to dialogues in need of repair as compared to human-human dialogues also in need of repair. Ultimately the data demonstrated that not only were there differences between "human-assistant" and "human-human" dialogue repair strategies, but that there were likewise differences among the assistants and the languages studied.

## I. Introduction

### Repair Initiators in Natural Language

Hayashi et al. (2018) characterizes human interaction by occurrences referred to as "infelicities, hitches and other troubles" which belong to the domain of behavior, named by Schegloff, Jefferson, and Sacks (1977), as "the organization of human repair". This organization systemetizes the practices through which participants mend miscommunications as they arise in exchanges. Interlocutors are furnished with resources to manage these disconnects in speaking, hearing, and understanding by way of a natural interactive system (Hayashi et al., 2018). These resources are the mechanisms by which humans are able to "repair" the communication between speakers when something fails (Dingemanse, et al., 2015).

Despite the vast variation in both grammar and meaning between languages, "they show a common infrastructure for social interaction which may be the universal bedrock upon which linguistic diversity rests" (Dingemanse, et al., 2015), that is to say - the phenomenon of communication repair is something universal in human communication. This proposition is supported by the presence of shared resources of repair initiation used in the same kind of contexts between unrelated languages. With this study, we aimed to systematically categorize how this phenomenon of natural language is processed by digital voice assistants which rely on an exchange of dialogue with natural and spontaneous human language.

### *Repair Initiators and Voice Assistants*

Voice Assistants (VAs) like Apple's Siri and Google's Google Assistant are virtual assistants that are able to perform tasks given by users by means of a voice command. The interactions of users with VAs has been shown to suffer from recurrent errors such as the assistant misunderstanding queries, accent/foreign languages, and contextual references of time or space (Motta and Quaresma, 2021). Voice assistant users might even recognize a phrase similar to "I'm sorry, I didn't hear that, let's try again..." (Cuadra et al., 2021). This demonstrates a type of response that is produced by assistants that serves to repair the interaction breakdowns that can occur when the assistant is unable to understand the user. The ubiquity of this phrase serves as an indication that voice assistants use at least one type of repair initiator for dialogue repairs.

## II. Previous Research
### *Repair Initiators - Humans*

Hayashi et al.'s aforementioned assertion that "repair is a part of a procedure of interaction" has been approached from different fields in the human and social sciences. For example, Goffman (1971) described an array of social behaviors which would become relevant later on for the analysis of communication repair; as the use of "I'm sorry" to initiate a repair, something that afterwards would be established as a common " use of an apology expression to justify a request for repetition" (Hayashi, 2018).

Nevertheless, Schegloff et al. (1977) findings were the ones to provide the starting point regarding repair studies. They noted key segments which were present in repair sequences: *the repairable*, *the repair initiator* and *the repair outcome*. As Cuadra (2021) reports, those authors also noticed that the repair can be done both from the speaker and the listener. This "self-repair" takes place, for example, whenever "the speaker notices a mistake before the listener has even the chance to respond". Cuadra also mentions that Schaefer's contributions helped to address the detection of repair segments with a model in which the speakers are "actively seeking evidence that they are being understood, and it is likely to initiate the repair when evidence of understanding is insufficient".

Repair strategies have been identified by multiple researchers and characterized in nine types. Schegloff, Jefferson, and Shacks in 1977 recognized five categories:

1. **Unspecified Repair** - In the unspecified type of repair there is a repair initiator 'huh' or 'what' – this type of strategy does not specify the trouble source, and these repair initiations usually yield a repetition of the trouble source turn.
2. **Interrogatives Repair** - A single question word such as 'who', 'where', or 'when' as repair initiations; this type of strategy specifies a trouble source of the prior turn.
3. **Partial Repeat Plus A Question Word** - there is a question word with the partial repeat of the trouble source turn.
4. **Partial Repeat Repair** - there is a partial repeat of the trouble source turn that is used for repair initiation.
5. **Understanding Check Repair** - the initiator is "you mean" plus a possible understanding of prior turn; here, the listener initiates the repair at the trouble source by giving an alternate understanding of the trouble source, then the speaker completes the repair on the next turn.

Later on, some studies dealt with repair techniques used by second language learning students (Egbert, 1998; Liebscher & Dailey-O'Cain, 2003) which differed from the ones aforementioned because those require skills that may not be yet developed:

6. **Request for Repetition** - shows a repetition of the trouble source (just as the native speaker forms of unspecified).
7. **Request for Definition** - the student would initiate the repair using the word that needs a translation of.
8. **Nonverbal Resources Repair** - yields a repetition of the trouble source turn or a modified version of original utterances.
9. **Correction Repair** - the teacher corrects the student using either verbal cues to indicate the student's utterance was wrong or by simply giving the student the correct word

To perform the comparison of the utterances from the voice assistants in this study, we created a new schema based in part on these nine types of repair techniques and strategies used by human speakers that all these previous researchers came up with (Schegloff 1977; Egbert, 1998; Liebscher & Dailey-O'Cain, 2003; Cho, 2010).

*Repair Initiators - Voice Assistants*

Given that humans program and train voice assistants, there is an underlying theme of errors and specific repair structures in social interactions with these conversational interfaces. Cuadra et al. (2021) highlighted that research into the methods that machines use to repair dialogues began with phone interaction systems. These systems gave rise to the previously mentioned "I'm sorry, I didn't hear that, let's try again..." phrase. Early phone dialogue systems tended to not repair the breakdowns that occured when the user failed to understand the system, with the researchers noting that users frequently responded to these types of situations by not speaking at all. Cuadra et al. also notes that repair strategies that simply reiterate or reformulate the original query of the user can help in getting the dialogue back on track, but can still be problematic if the line of inquiry or dialogue is incorrect.

Dialogues with VAs that require repair are likewise plagued by problems regarding interactional language. Cuadra et al. describes research that demonstrates that users are less likely to initiate repairs with voice assistants or systems that are not represented by humans in general due to intersubjectivity in which one party thinks that the other party knows what their point of view is. Thus the user person does not initiate a repair as they believe that the assistant cannot understand their repairs. Likewise, they point out that studies that used chatbots to repair dialogue among other chatbots found that these bots had difficulties understanding the points of view of the other bots.

Motta and Quaresma's contemporary study conducted in 2021, *Users' Error Recovery Strategies in the Interaction with Voice Assistants,* identified eight distinct strategies that Google Assistant and Siri along with their corresponding frequencies of use:

1. **Different task** (n = 40)
    a. The VA perform an activity or task other than the task required

2. **Wrong information** (n=33)

   a. The VA adds the wrong details to the appointment

3. **Input failure** (n = 27)
   a. The VA does not capture any part of the user's command

4. **Interruption** (n = 19)
   a. The VA stops to capture input halfway through users' command

5. **Misrecognition** (n = 12)
   a. The VA misrecognized one or more words in the input

6. **Request for manual interaction** (n=9)
   a. The VA asks the user to interact manually to add information, save, or cancel events. This was considered an error type as users consider requests for manual interaction as a nuisance (3,9)

7. **Error messages** (n=11)
   a. The VA explicitly tells the user that an error has occurred (e.g., "Sorry, I didn't understand") or that it cannot perform the task

8. **Instructions** (n = 178)
   a. The VA explicitly tells the user how to proceed in case of error or offer directions to move the interaction forwards (e.g., "Say 'yes' to save"). Questions (e.g., "What is the event's title?") and confirmations (e.g., "I created the event") were considered indirect instructions as they provide cues to advance interactions. We considered for analysis instructions that were provided after or along with an error or that preceded changes of approach

### III. Problem Statement

Interactions in every language require dialogue repair after instances of mishearing or in cases in which clarification is needed. In a similar vein, interaction with virtual voice assistants mimics human dialogue, but has unique obstacles to overcome in order to successfully emulate a convincing interlocutor. Our research aims to compare conversations containing dialogue repair turns between humans and similar conversations between humans and voice assistants to examine how these assistants' repair strategies deviate (if at all) from person to person interactions.

Repair initiators such as 'huh' serve to mend a fundamental disconnect between interlocutors in a conversation. Most research that explores this phenomenon has focused on human to human interaction. We propose, however, an examination of dialogue repair structure within a virtual context wherein the dialogue initiator is human and the party that initiates the repair - after a misunderstanding, for example - is a virtual voice assistant.

Our research entails a comparative analysis of the use of repair initiators in human-human interactions and human-assistant interactions. This analysis looks at the differences between utterances (in both English and Spanish) used as repair initiators by virtual assistants (Google Assistant and Apple's Siri) and those used by humans, and the differences in responses elicited by the human dialogue initiator to the presence of these repair initiators.

### IV. *Hypotheses*

The research at hand aims to respond to two capacities of voice assistants: the reaction and the production of repair initiators as 'huh' in English ('eh' in Spanish). We predict that Google Assistant and Apple's Siri, when provoked by utterances that are unintelligible, will not produce discourse repair structures such as "huh" (as seen in Digamanse, 2013), rather they respond in a way that is generalized for clarity and universal understanding in both Spanish and English – that is, producing utterances such as, "I am sorry, I could not hear you." or "Could you repeat, please?". In the interest of understanding the user at any cost, however, we believe the assistants will respond when the user is the one eliciting the dialogue repair using words of any register, such as "huh".

### V. *Methodology*

To test the capability of the voice assistants to produce and react to repair initiators, two tests were devised. The first test ("Task 1") focused on documenting the production of repair initiators by the assistants:

1. Initiate a dialogue with Google Assistant and Apple Siri in both English and Spanish using the keyword for each (n=30; 15 in English, and 15 in Spanish).
2. When prompted, begin a request but end it with an intentionally unintelligible phrase.
3. Allow the assistant to respond.
4. Record the response of the assistant.

The second test ("Task 2") intended to document how the assistants would react to repair initiators on behalf of the user:

1. Initiate a dialogue with Google Assistant and Apple Siri in both English and Spanish using the keyword for each (n=20; 10 in English, and 10 in Spanish).
2. When prompted, begin a request that requires extra information to be requested by the assistant, such as "Set a reminder".
3. Allow the assistant to request the additional information.
4. Respond to the request with different solicitations for clarification: 'huh?' and '¿eh?'.
5. Record the response of the assistant.

Finally, a third task ("Task 3") sought to compare human-assistant dialogues to human-human dialogues. In order to be able to analyse the data, the responses from the VAs were divided into 10 categories and sorted results accordingly. This task was directly addressed within the discussion portion of this paper.

## VI.   Results

Employing the methodology listed above, the data presented in the appendices listed at the end of this paper was obtained. From this data – and taking influence from the previous repair strategy categories created research from Schegloff, Egbert, Liebscher & Dailey-O'Cain, Cho, and Motta and Quaresma – we synthesized a list which categorized the VAs' techniques (i.e. strategies) used to repair discourse. This compendium consists of ten strategies that are presented here with examples of utterances obtained which demonstrated the corresponding strategy:

1. **Attempts to fulfill requests given the information, excluding the unintelligible part.**
   a. User: Hey Siri, I have an appointment at the dentist for [unintelligible phrase].
      Siri: I didn't find any appointment at "dentist" in the next three months.

2. **Takes a literal interpretation of the unintelligible part of the utterance.**
   a. User: Hey Siri, how can I get to the street ["blah blah"]?
      Siri: I could not find any places matching "Street blah blah".

   b. User: Ok Google, send a text message to Adrian and Sofia that says the following: [unintelligible phrase].
      Google: [Creates text that says "the following" and asks for confirmation to send].

3. **Fails to understand or mishears - sometimes proposes an error, may be accompanied by the system asking for the user to ask again.**
   a. User: Hey Siri, play the song [unintelligible sentence].
      Siri: Uh oh, there's a problem. Please, try again.

   b. User: Hey Siri, [unintelligible phrase].
      Siri: I'm not sure I understand.

4. **Not capable of fulfilling requests (sometimes cancels the request due to this inability).**
   a. User: Hey Siri, how long does it take from Balmes street to [unintelligible phrase]?
      Siri: Sorry, I can't give you traffic to that location.

5. **Asks for the appropriate information that occupies the space of the unintelligible speech.**
   a. User: Hey Siri, call [unintelligible phrase].
      Siri: Who would you like to call?

6. **Searches the internet for information that it attempts to parse from the unintelligible part of the phrase.**
    a. User: Hey Siri, what is the capital of [unintelligible phrase]?
       Siri: I found this on the web.

7. **Uses information stored about the user to fulfill requests.**
    a. User: Ok Google, how can I get to the street [unintelligible phrase]?
       Google: The best way to get to [user's home address] is…

8. **Opens (or proposes) an application that the user can further use to ask the request again to better fulfill it.**
    a. User: Ok Google, play the new album of [unintelligible phrase].
       Google: [Opens YouTube app].

9. **Gives the user a list of instructions (or the website to find the information) for the operation of the assistant.**
    a. User: Ok Google, [unintelligible phrase].
       Google: [Gives instructions about possible actions with the assistant].

10. **Attempts to phonetically transcribe the unintelligible part of the utterance.**
    a. User: Ok Google, [unintelligible phrase].
       Google: [no response, unintelligible phrase partially parsed].

We then used these ten strategies to categorize the responses elicited from the voice assistants to perform the following quantitative analysis.

## **Task 1**

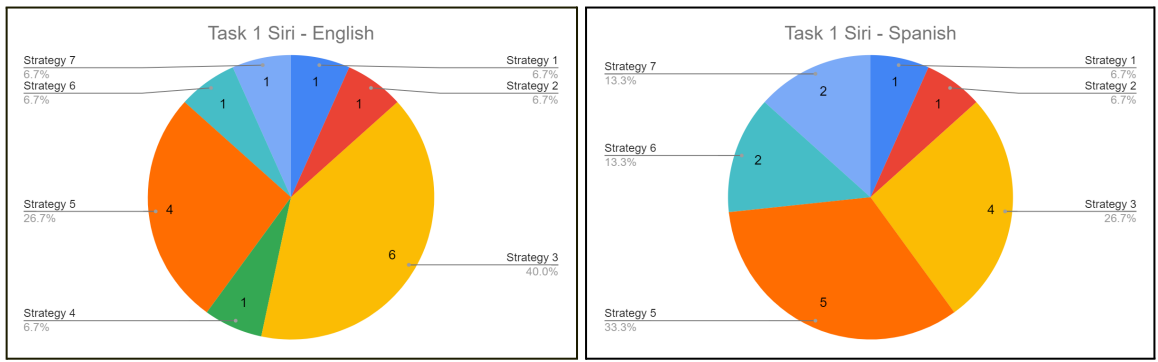

*Figure 1*. Task 1 - Siri: English and Spanish

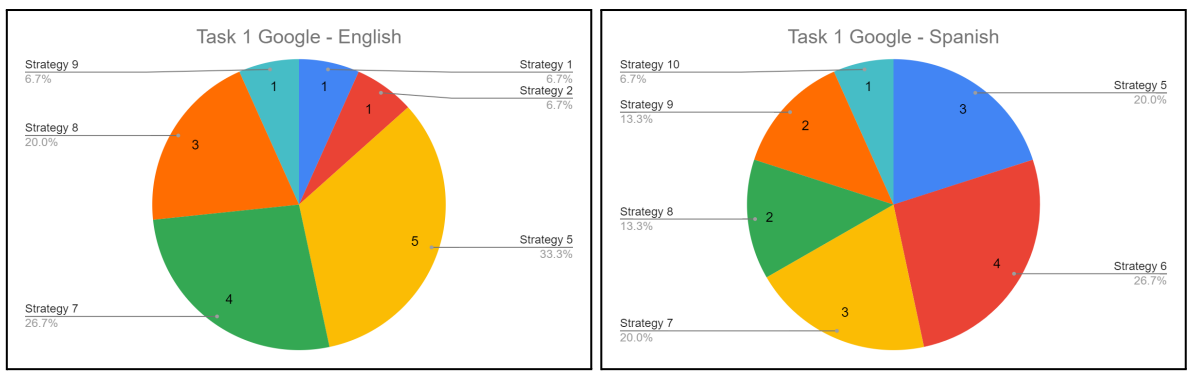

*Figure 2*. Task 1 - Google Assistant: English and Spanish

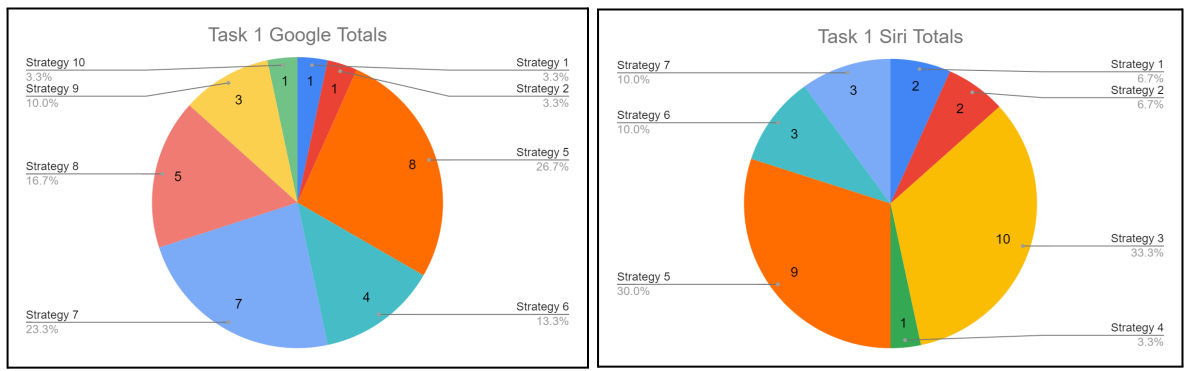

*Figure 3*. Task 1 - Google Assistant vs Siri Totals: English and Spanish

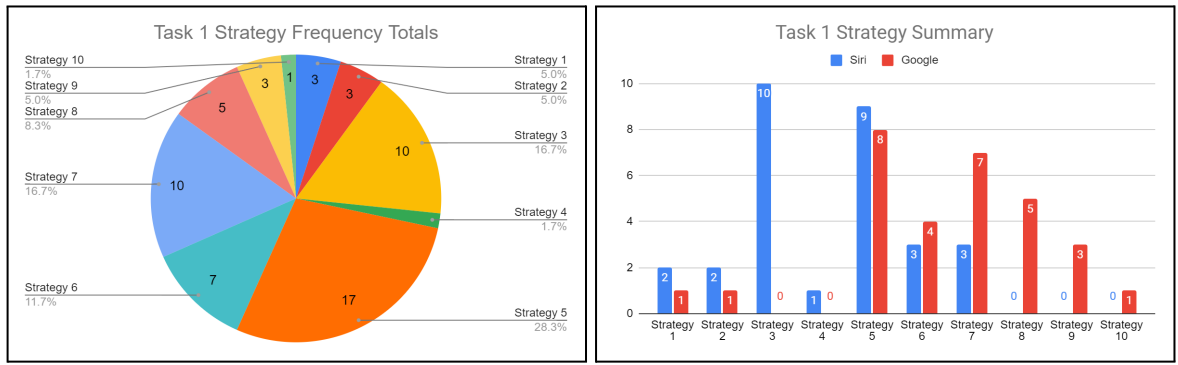

*Figure 4*. Task 1 - Strategy Summary

## Task 2

*Note: Not all results total to ten questions due to Siri giving more answers than Google in some instances.*

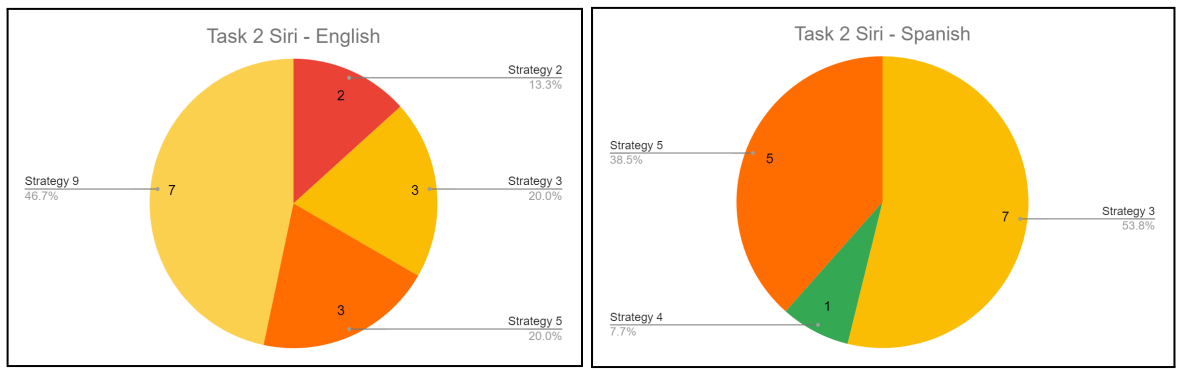

*Figure 5*. Task 2 - Siri: English and Spanish

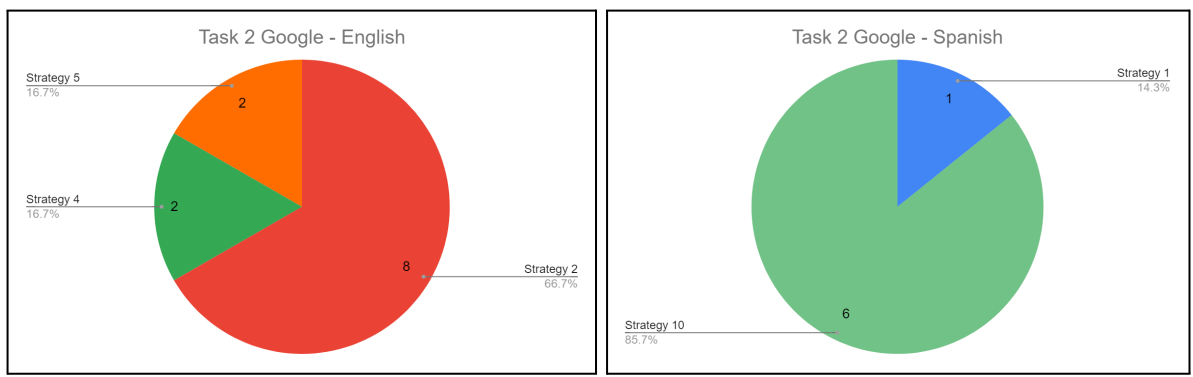

*Figure 6*. Task 2 - Google Assistant: English and Spanish

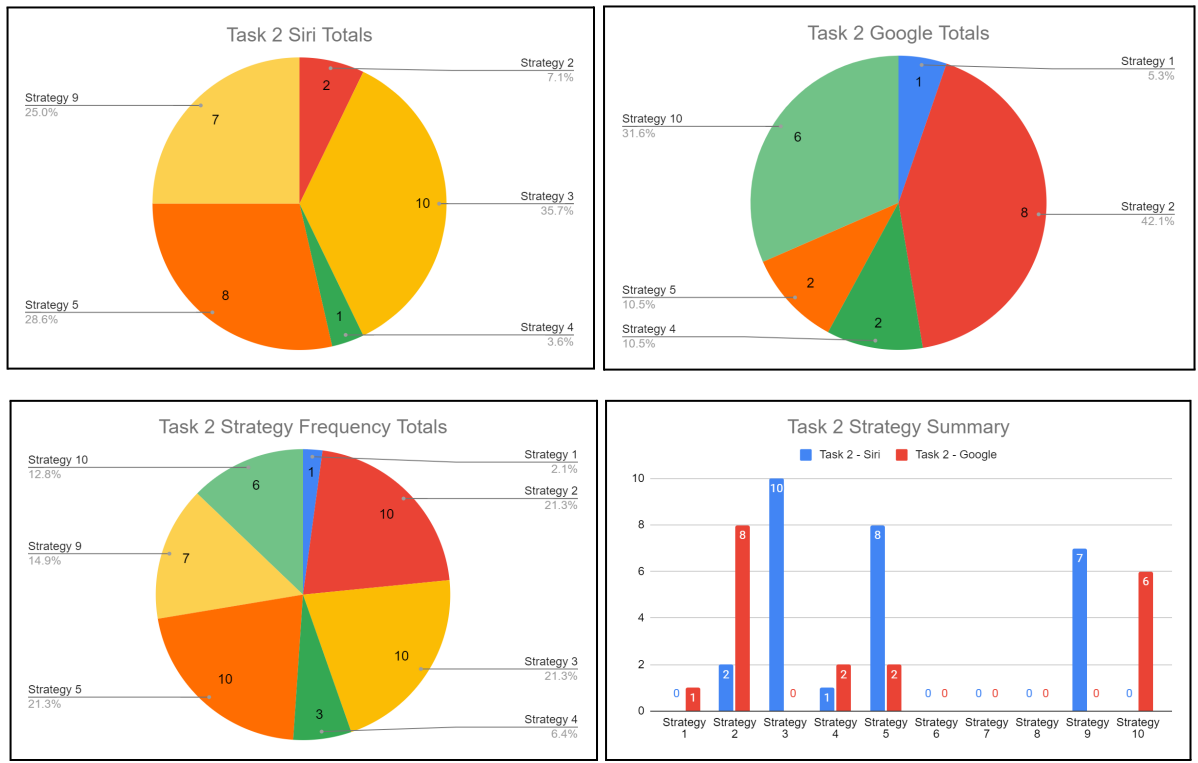

*Figure 7*. Task 1 and 2 - Strategy Summaries

## VII. Discussion

### Task 1

Task 1 involved the user producing an unintelligible phrase in order to provoke a repair response from the assistant. This exercise of introducing malformed queries produced differing responses for each assistant.

Google Assistant's English implementation tended to use strategies 5, 7, and 8 (asking for the appropriate information, using information stored about the user to fulfill requests, and opening an application, respectively). All of these are what might be considered to be "positive feedback" in that they are all attempts at repairing conversation in one way or the other. Google Assistant also lacked strategies 3, 4, 6, and 10 (fails to understand or mishears, not capable of fulfilling request, searches the internet for information, and attempts to transcribe, respectively), many of which could be considered "negative feedback" repair techniques which either do not initiate a repair at all or weakly attempts a repair.

In the Spanish task Google's Assistant produced strategies 5 and 7 the most – overlapping with the English task's most used strategies. Meanwhile, this task lacked strategies 1, 2, 3, and 4. This mimics the attempt at positive feedback while avoiding negative feedback of the English task.

Siri in English for this task tended to use strategies 6 (searches the internet for information that it attempts to parse from the unintelligible part of the phrase) and 4 (not capable of fulfilling request – sometimes cancels the request due to this inability) in near equal instances (10 instances vs. 9 instances). While strategy 6 would be considered positive feedback, strategy 4 would not as it is unable to provide a solution. Siri also lacks strategies 8, 9, and 10. all of which would be considered as positive feedback. In the Spanish task Siri focuses on the use of positive feedback strategy 5, similar to Google's Assistant.

In summation for this task, the most used strategy is 5 – a positive feedback strategy. The next most used strategies are 3 and 7, with 3 – a negative feedback strategy – being attributed completely to Siri and the majority of 7 – a positive feedback strategy – nearly completely attributable to Google Assistant.

### Task 2

Task 2 involved the user initiating a dialogue with an assistant using a request that requires additional information such as "set a reminder". The assistant would then request the additional information and the user would respond to the request with the dialogue repair initiator *huh?* (English) or *¿eh?* (Spanish). The assistant's response was then recorded. Of note in this task is that although ten questions were asked of each assistant in both English and Spanish, not all questions were answered - that is, Google assistant would occasionally not answer to the user's repair initiator at all.

This task sees Google Assistant in English mainly using strategy 2 and to a lesser extent strategies 4 and 5 for a total of 3 possible strategies. While strategy 2 (taking a literal interpretation of the unintelligible part of the utterance) is still considered a positive strategy, in comparison to something like strategy 5 it could be considered of lesser quality as the former almost always produces undesirable results.

Google Assistant in Spanish fairs even worse. The most used strategy is 10 (attempts to phonetically transcribe the unintelligible part of the utterance) a positive feedback strategy though a poor one, poorer even than 2 from the English task. This strategy almost never produces a useful resolution for the user. The only other strategy seen used was strategy 1 – a positive strategy which occurred only once. The assistant in this language task even ended conversations before the repair initiator could be used, for example: User: Pon una alarma.

Google Assistant: [Gives instructions to start an alarm]. In this example the assistant fails to advance the dialogue far enough to necessitate the repair initiator.

Apple's Siri likewise performed worse in this task than in task 1. The most used strategy was strategy 9 which gives the user a list of instructions (or the website to find the information) for the operation of the assistant, a response the user would typically not prefer, making it negative feedback. Siri similarly lacked the use of many strategies, only having 4 of 10, though this was more than Google's assistant. In Spanish Siri tended to use strategy 3 (failing to understand or mishears) – a negative feedback strategy. It did, however, have more instances of positive feedback and more resources in general than Google Assistant.

From this we can see that Task 2 produced different strategies than Task 1. Overall the strategies 2, 3, and 5 were most used, with only one of these being positive feedback that progressed the dialogue.

**Task 1 and 2 Overview**

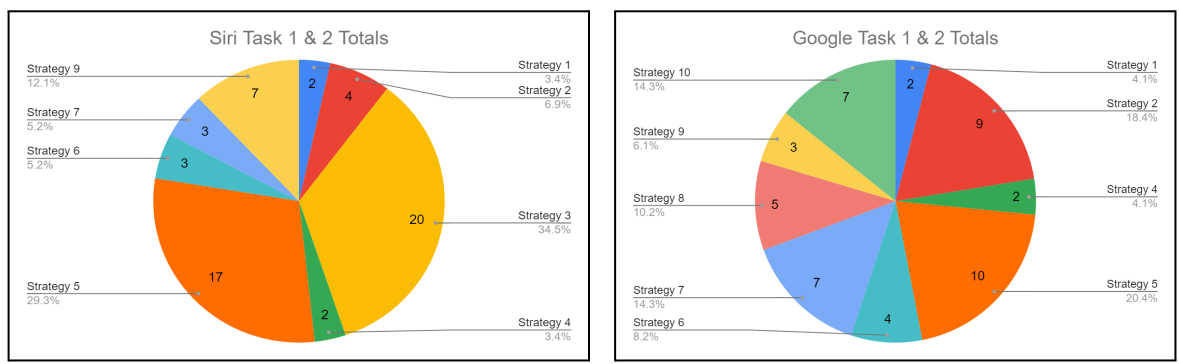

*Figure 8. Task 1 and 2 - Strategy Totals per Assistant*

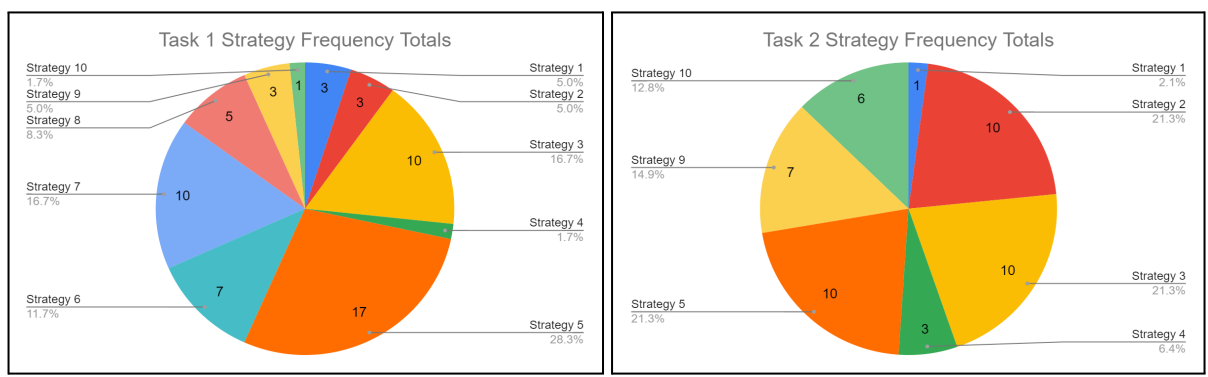

*Figure 9. Task 1 and 2 - Strategy Frequency Totals*

Looking at the strategies used in Task 1 we see an overall pattern emerging. The most most used strategies were 5, 7, and 3 with the first two strategies being positive feedback and the final one being negative feedback. As noted before, strategy 3 was exclusively used by Siri. Likewise of note is that within this task all strategies used at some point, with some being used much less than others. Otherwise, Task 2 showed strategies 2, 3, and 5 emerging as the most used, with only one being a positive feedback response. This task had less overall strategies possible with a more equal usage of those that were usable. This, coupled with the lack of positive feedback responses, indicated a lack of acceptable responses and a lack of focus on the part of the assistants regarding the overall goal.

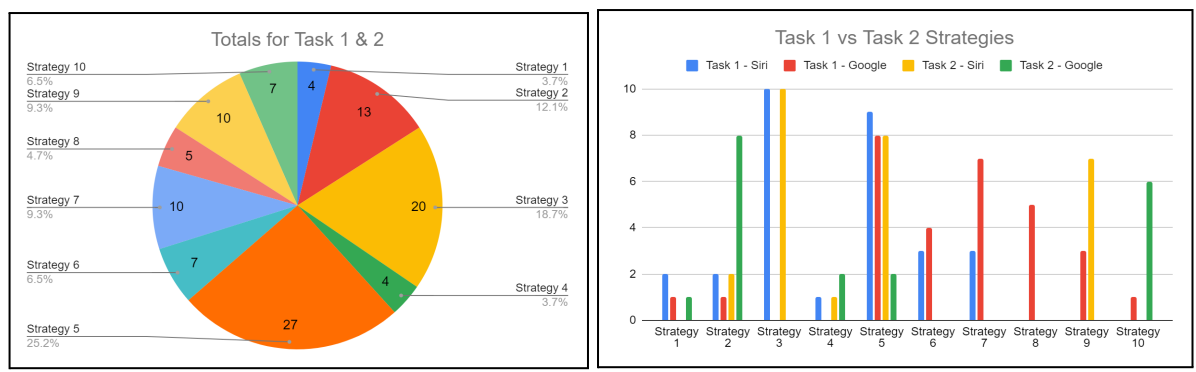

*Figure 10.* Task 1 and 2 Strategy Total Combined

Overall, certain strategies emerged as the forerunners when looking at all of the strategies both machines used. The number one most used was strategy 5 (accounting for around 25% of all responses) which upon asking for more information seeks to disambiguate what the user said. Users of assistants are most likely familiar with this utterance as a response. A negative strategy. strategy 3, occupied the second most used spot – as shown beforehand, this was exclusively used by Apple's Siri. Given this, we have observed a distinct difference in philosophy regarding the goal of the assistants: Apple's Siri accepts mistakes and readily uses strategies of negative feedback such as strategy 3. Meanwhile Google Assistant aims to avoid a negative feedback strategy at all cost – it will use many strategies, even poor quality ones such as transcribing the unintelligible phrases it hears, in order to progress the dialogue along in any meaningful way. This difference also extends to performance of the assistants in the two languages that were assessed, wherein we saw an overall higher level of development in the English tasks, possibly due to the resources provided to developers for English tasks, which in turn produces a more well rounded assistant in the language.

## **Task 3**

From the data collected, it becomes clear that human and machine strategies differ in many ways and overlap in very few areas. From what we have observed the most common positive feedback repair strategy was the one shared with human-human interactions – strategy 5 which seeks to disambiguate what was said. This strategy shares characteristics with the 3rd and 4th core repair techniques used between humans as described by Schegloff et al. (1977):

> 3. *Partial Repeat Plus A Question Word Repair Example*
>
> Speaker 1: Was last night the first time you met Missus Kelly?
> **Speaker 2: Met whom?**
> Speaker 1: Missus Kelly,
>
> 4. *Partial Repeat Repair Example*
>
> Speaker 1: Well Monday, lemme think. Monday, Wednesday, an'
> Fridays I'm home by one ten.
> **Speaker 2: One ten?**
> Speaker 1: Two o'clock. My class ends one ten.

These human strategies, however, display a level of informality that voice assistants possibly avoid for both clarity and politeness reasons.

The second most used strategy in machines, strategy 3 which fails to understand and causes the system to ask the user to ask for the strategy, mirrors human-human strategy 6 seen in second language learning environments that asks for repetition of the misheard utterance:

> 6. *Request for Repetition Example*
>
> Teacher: Was hat Ihnen dieses Semester im Deutschkurs nicht gefallen
> (What, did you not like this semester in your German course?)
> **Student: Noch einmal? (Once more?)**
> Teacher: Was hat Ihnen dieses Semester im Deutschkurs nicht gefallen
> (What, did you not like this semester in your German course?)

These types of repair initiators tend to focus on the direct pursuit of knowledge rather than something like an informal, human-human 'huh' – machines, similar to SLA learners, focus on making their repair initiators clear and direct in order to elicit the exact information that they need.

## VIII.     Conclusion

From the results of this study, we have demonstrated the contrast between English and Spanish repair strategies from Siri and Google Assistant in two tasks. English tasks appear to be more proficient in both the production and the reaction to repair initiators; a fact that could be due to the more time, resources and research invested in it.

While documenting the possible repair initiators by the assistant in response to an unintelligible phrase, differences arose between Google Assistant and Siri in both Spanish and English. Generally, Google Assistant, when prompted with an unintelligible phrase, either tends to ask for an appropriate answer or uses stored information about the user. However, Siri tends to behave differently in these contexts and usually fails to understand or mishears the sentences prompted by the user though in some cases it also asks for appropriate information or tries to provide the user something. Ultimately, this provides an impression that Google Assistant is more focused on repairing discourse at any cost (i.e. using strategies we deemed 'positive feedback strategies') , while Siri readily displays errors when a repair is needed – that is, it more frequently uses 'negative feedback strategies'.

Data gathered about how assistants reacted to repair initiators (English 'huh'/ Spanish 'eh') on behalf of the user also showed differences in the behavior of the two assistants and between languages. Regarding the reactions obtained with Siri tended to fail to understand or mishear, ask for the appropriate information that occupies the unintelligible speech, or give instruction for the operation of the assistant. Google Assistant mostly took a literal interpretation of the unintelligible part of the utterance or attempted to phonetically transcribe it.

We have seen how humans and machines differ greatly. The most commonly used strategy by machines is asking for the appropriate information that occupies the space of the unintelligible speech, and this is similar to repair techniques we have seen in humans – namely the partial repetition of an utterance while adding a question word, or even partial repetition repair. However, possibly due to the perceived informality of some repair strategies, the use differed between machines and humans with machines tending to avoid phrases such as 'huh' or 'what?'.

The second strategy that machines performed the most was the failure to understand or the mishearing followed by the petition to the user to repeat the request; which could be associated with the request for repetition strategy performed by humans (and, language learners, above all). Hence, as we saw, machines mostly confine themselves to SLA techniques like requests for repetition which, like the asking for appropriate information shown above, is ostensibly due to the pursuit of expressing formality and direct knowledge , unlike more vague repair strategies like 'huh' in humans.

With this in mind, the hypotheses proposed were only partially confirmed. On the one hand, we have seen that it is true that Google Assistant and Apple's Siri, when provoked by utterances that are unintelligible, do not produce discourse repair structures such as "huh?" or "¿eh?". Also, as we presumed, they sometimes respond in a way that is generalized for clarity and universal understanding in both Spanish and English. However, our second assumption that the assistants would respond when the user elicited dialogue repair using words such as "huh?" or "¿eh? was only validated in some of the phrases recorded.

*Appendices*

**Appendix 1.**

Task 1

**A. Questions and Answers from Siri in English:**

1. User: Hey Siri, I have an appointment at the dentist for [unintelligible phrase].
   Siri: I didn't find any appointment at "dentist" in the next three months.

2. User: Hey Siri, how can I get to the street ["blah blah"]?
   Siri: I could not find any places matching "Street blah blah".

3. User: Hey Siri, play the song [unintelligible sentence].
   Siri: Uh oh, there's a problem. Please, try again.

4. User: Hey Siri, [unintelligible phrase].
   Siri: I'm not sure I understand.

5. User: Hey Siri, are there any [unintelligible phrase] in Barcelona?
   Siri: I didn't get that. Could you try again?

6. User: Hey Siri, how long does it take from Balmes street to [unintelligible phrase]?
   Siri: Sorry, I can't give you traffic to that location.

7. User: Hey Siri, call [unintelligible phrase].
   Siri: Who would you like to call?

8. User: Hey Siri, send a message to [unintelligible phrase].
   Siri: Who do you want to send it to?

9. User: Hey Siri, what's the forecast for today in [unintelligible phrase].
   Siri: It's currently partly cloudy and nine degrees.

10. User: Hey Siri, where's a [unintelligible phrase] nearby?
    Siri: I didn't get that, could you try again?

11. User: Hey Siri, play the new album of [unintelligible phrase].
    Siri: Sorry, something's gone wrong. Please try again.

12. User: Hey Siri, send a text message to Maria that says: [unintelligible phrase].
    Siri: What do you want to say?

13. User: Hey Siri, create a reminder for [unintelligible phrase].
    Siri: What do you want to be reminded about?

14. User: Hey Siri, what is the capital of [unintelligible phrase]?
    Siri: I found this on the web.

15. User: Hey Siri, how long is the [unintelligible phrase]?
    Siri: I didn't get that. Could you try again?

**B. Questions and Answers from Siri in Spanish:**

1. User: Oye Siri, tengo una cita en el dentista para [unintelligible phrase].
   Siri: ¿Para qué día es la cita?

2. User: Oye Siri, ¿cómo puedo llegar a la calle ["blablablá"]?
   Siri: Lo siento, no he encontrado calle blablablá

3. User: Oye Siri, pon la canción [unintelligible phrase].
   Siri: Lo siento, [user name], no sé qué quieres que cambie exactamente.

4. User: Oye Siri, [unintelligible phrase].
   Siri: Lo siento, no te he oído bien. ¿Puedes repetirme lo que has dicho?

5. User: Oye Siri, ¿hay algún [unintelligible phrase] en Barcelona?
   Siri: Buena pregunta. / Creo que no te entiendo.

6. User; Oye Siri, ¿cuánto se tarda desde la calle Balmes a [unintelligible phrase]?
   Siri: El tráfico desde Carrer de Balmes, Barcelona, hasta tu ubicación actual es normal, así que calculo que tardarás 14 minutos por C-31.

7. User; Oye Siri, llama a [unintelligible phrase].
   Siri: ¿A quién quieres llamar?

8. User: Oye Siri, envía un mensaje a [unintelligible phrase].
   Siri: ¿A quién quieres enviárselo?

9. User: Oye Siri, ¿cuál es el pronóstico de hoy en [unintelligible phrase]?
   Siri: Ahora mismo hay algunas nubes y la temperatura es de nueve grados

10. User: Oye Siri, ¿dónde hay una [unintelligible phrase] cerca?
    Siri: ¿Qué tipo de negocio estás buscando? / No he encontrado sitios con estas características

11. User: Oye Siri, reproduce el nuevo álbum de [unintelligible phrase].
    Siri: Uy, ha habido un problema. Vuelve a intentarlo.

12. User: Oye Siri, envía un mensaje de texto a María que diga: [unintelligible phrase].
    Siri: ¿Qué quieres que diga?

13. User: Oye Siri, crea un recordatorio para [unintelligible phrase].
    Siri: ¿Qué quieres que te recuerde?

14. User: Oye Siri, ¿cuál es la capital de [unintelligible phrase]?
    Siri: Vale, esto es lo que he encontrado en Internet sobre "¿Cuál es la capital de?". Échale un vistazo.

15. User: Oye Siri, ¿cuánto mide el [unintelligible phrase]?
    Siri: Esto es lo que he encontrado en Internet.

**C. Questions and Answers from Google Assistant in English:**

1. User: Ok Google, set an appointment for the dentist at [unintelligible phrase].
   Google: Alright, when is the event?

2. User: Ok Google, how can I get to the street [unintelligible phrase]?
   Google: The best way to get to [user's home address] is…

3. User: Ok Google, play the song [unintelligible sentence].
   Google: Opening YouTube

4. User: Ok Google, [unintelligible phrase].
   Google: [Gives instructions about possible actions with the assistant].

5. User: Ok Google, [unintelligible sentence].
   Google: (Presents options of possible requests)

6. User: Ok Google, are there any [unintelligible phrase] in Barcelona?
   Google: According to Wikipedia Camp Nou is Barcelona stadium in Spain

7. User: Ok Google, how long does it take from Balmes street to [unintelligible phrase]?
   Google: [Tells time to get to Blames street from home location]

8. User: Ok Google, call [unintelligible phrase].
   Google: Who do you want to call?

9. User: Ok Google, send a message to [unintelligible phrase].
   Google: Which one do you want to use? [Presents installed apps used for texting].

10. User: Ok Google, what's the forecast for today in [unintelligible phrase].
    Google: [Gives forecast for user's current location].

11. User: Ok Google, where's a [unintelligible phrase] nearby?
    Google: [Gives information about user's current location].

12. User: Ok Google, play the new album of [unintelligible phrase].
    Google: [Opens YouTube app].

13. User: Ok Google, send a text message to Adrian and Sofia that says the following: [unintelligible phrase].
    Google: [Creates text that says "the following" and asks for confirmation to send].

14. User: Ok Google, create a reminder for [unintelligible phrase].
    Google: What's the reminder?

15. User: Ok Google, what is the capital of [unintelligible phrase]?
    Google: What is the capital of what?

16. User: Ok Google, how long is the [unintelligible phrase]?
    Google: How long is what?

**D. Questions and Answers from Google Assistant in Spanish:**

1. User: Ok Google, apunta que tengo una cita en el dentista para [unintelligible phrase].
   Google: Vale, ¿cuándo es el evento?

2. User: Ok Google, ¿cómo puedo llegar a la calle [unintelligible phrase]?
   Google: Aquí tienes unos resultados [searches Google for the word "calle"].

3. User: Ok Google, pon la canción [unintelligible phrase].
   Google: [Opens YouTube app].

4. User: Ok Google, [unintelligible phrase].
   Google: [no response, unintelligible phrase partially parsed].

5. User: Ok Google, ¿hay algún [unintelligible phrase] en Barcelona?
   Google: [Searches for things to do in Barcelona on Google].

6. User: Ok Google, ¿cuánto se tarda desde la calle Balmes a [unintelligible phrase]?
   Google: [Gives time it takes to get from user's home location to Balmes street].

7. User: Ok Google, llama a [unintelligible phrase].
   Google: ¿A quién quieres llamar?

8. User: Ok Google, envía un mensaje a [unintelligible phrase].
   Google: [Searches Google for how to use the send a message feature on assistant].

9. User: Ok Google, ¿cuál es el pronóstico de hoy en [unintelligible phrase]?
   Google: [Gives user's current location forecast].

10. User: Ok Google, ¿dónde hay una [unintelligible phrase] cerca?
    Google: [Gives map of user's current location].

11. User: Ok Google, reproduce el nuevo álbum de [unintelligible phrase].
    Google: [Opens YouTube app].

12. User: Ok Google, envía un mensaje de texto a Adrian y Sofia que diga lo siguiente: [unintelligible phrase].
    Google: [Gives instructions on how to ask for assistant to send message, asks for more information such as content of message].

13. User: Ok Google, crea un recordatorio para [unintelligible phrase].
    Google: ¿Cuál es el recordatorio?

14. User: Ok Google, ¿cuál es la capital de [unintelligible phrase]?

    Google: [Searches Google for the word capital].

15. User: Ok Google, ¿cuánto mide el [unintelligible phrase]?

    Google: [Searches Google for the phrase ¿cuánto mide el…?].

**Appendix 2.**

## Task 2

**A. Questions and Answers from Siri in English:**

1. User: Set an alarm.
   Siri: Set the alarm for when?
   User: huh?
   Siri: I'm here to help, get to know Siri at Apple.com.

2. User: Set a reminder.
   Siri: What do you want to be reminded about?
   User: huh?
   Siri: Ok. Added. [Adds "huh" as a reminder].

3. User: Set a timer.
   Siri: For how long?
   User: huh?
   Siri: I'm here to help, get to know Siri at Apple.com

4. User: Add something to my calendar.
   Siri: What date and time is your appointment?
   User: huh?
   Siri: I'm here to help, get to know Siri at Apple.com / What date and time is your appointment? / Uh oh, I missed what you said. There's a problem. Please try again.

5. User: Send a text message.

    Siri: Who do you want to send it to?

    User: huh?

    Siri: I'm here to help, get to know Siri at Apple.com. / Who do you want to send it to? / I didn't get that. Could you try again?

6. User: Add to my shopping list.

    Siri: I didn't find a shopping list. Do you want to create one?

    User: huh?

    Siri: What do you want to be reminded about?

7. User: Make a call.

    Siri: Who would you like to call?

    User: huh?

    Siri: I'm here to help, get to know Siri at Apple.com.

8. User: Start a video chat.

    Siri: Who would you like to FaceTime with?

    User: huh?

    Siri: I'm here to help, get to know Siri at Apple.com.

9. User: Start a list.

    Siri: Sorry I can't help with that. Let me know if you want me to open the Reminders App.

    User: huh?

    Siri: I'm here to help, get to know Siri at Apple.com.

10. User: Send an email.

    Siri: To whom shall I send it?

    User: huh?

    Siri: I couldn't find "huh" in your contacts. To whom shall I send it?

**B. Questions and Answers from Siri in Spanish:**

1. User: Pon una alarma.
   Siri: ¿Para qué hora?
   User: ¿eh?
   Siri: No te he entendido. ¿Me lo puedes repetir?

2. User: Configura un recordatorio.
   Siri: ¿Qué quieres que te recuerde?
   User: ¿eh?
   Siri: Lo siento, no he entendido bien lo que has dicho. ¿Qué quieres que te recuerde?

3. User: Pon un temporizador.
   Siri: ¿De cuánto tiempo?
   User: ¿eh?
   Siri: ¿De cuánto tiempo? / ¿De cuánto tiempo?

4. User: Crea un evento de calendario.
   Siri: ¿Cuál es la fecha y la hora de esta cita?
   User: ¿eh?
   Siri: Lo siento, no te he entendido. ¿Podrías repetirlo? / ¿Cuál es la fecha y la hora de esta cita?

5. User; Envía un mensaje de texto.
   Siri: ¿A quién quieres enviarselo?
   User: ¿eh?
   Siri: Lo siento, no he entendido bien. ¿Puedes repetirme lo que has dicho?

6. User: Agrega algo a mi lista de compras.
   Siri: No he encontrado ninguna lista llamada "Compras". ¿Quieres que la cree?
   User: ¿eh?
   Siri: Lo siento, no te he oído. ¿Podrías volver a decírmelo?

7. User: Empieza una llamada.
   Siri: ¿A quién quieres llamar?
   User: ¿eh?
   Siri: Lo siento, no te he oído bien. ¿Puedes repetirme lo que has dicho?

8. User: Empieza una llamada de vídeo.
   Siri: ¿A quién quieres llamar por FaceTime?
   User: ¿eh?
   Siri: Lo siento, no te he oído bien. ¿Puedes repetirme lo que has dicho?

9. User: Crea un listado.
   Siri: Lo siento, no puedo ayudarte con eso. Dime si quieres que abra la App "Recordatorios".
   User: ¿eh?
   Siri: Lo siento, no puedo ayudarte con eso.

10. User: Envía un correo electrónico.
    Siri: ¿A quién se lo mando?
    User: ¿Qué?
    Siri: ¿A quién se lo mando? / ¿A quién quieres enviarle el correo? / No te he entendido ¿Me lo puedes repetir?

**C. Questions and Answers from Google Assistant in English:**

1. User: Set an alarm.
   Google: Ok, when's the alarm?
   User: huh?
   Google: No problem, no new alarm set.

2. User; Set a reminder.
   Google: Ok, what's the reminder?
   User: huh?
   Google: [Sets reminder titled "huh"].

3. User: Set a timer.

    Google: Ok, for how long?

    User: huh

    Google: Sorry, for how long? / Sorry, I'm still not sure how long… / I'm still not sure what you mean so I canceled the timer.

4. User: Add something to my calendar.

    Google: Ok, what's the title of the event?

    User: huh?

    Google: [Adds event titled "huh"].

5. User: Send a text message.

    Google: Ok, who do you want to text?

    User: huh?

    Google: [Attempts to parse contacts similar to "huh"].

6. User: Add to my shopping list.

    Google: Ok, what do you want to add?

    User: huh?

    Google: Alright I added huh.

7. User: Make a call.

    Google: Ok, who do you want to call?

    User: huh?

    Google: [Attempts to parse contacts similar to "huh"].

8. User: Start a video chat.

    Google: Ok, who do you want to video call?

    User: huh?

    Google: [Attempts to parse contacts similar to "huh"].

9. User: Start a list.

    Google: Ok, what do you want to call it?

    User: huh?

    Google: [Creates a list called "huh"].

10. User: Send an email.

    Google: Ok, who do you want to email?

    User: huh?

    Google: Looks like there's no email with that name.

**D. Questions and Answers from Google Assistant in Spanish:**

1. User: Pon una alarma.

    Google: [Gives instructions to start an alarm].

2. User; Configura un recordatorio.

    Google: ¿Cuál es el recordatorio?

    User: ¿eh?

    Google: [No response, unintelligible phrase partially parsed].

3. User: Pon un temporizador.

    Google: [Gives instructions to set a timer].

4. User; Agrega algo a mi calendario.

    Google: ¿Cómo quieres que se llame el evento?

    User: ¿eh?

    Google: [No response, unintelligible phrase partially parsed].

5. User: Manda un mensaje de texto.

    Google: ¿A quién?

    User: ¿eh?

    Google: [No response, unintelligible phrase partially parsed].

6. User: Agrega algo a mi lista de compras.

    Google: ¿Qué quieres añadir?

    User: ¿eh?

    Google: [Displays list called "compras"].

7. User: Empieza una llamada.

    Google: ¿A quién quieres llamar?

    User: ¿eh?

    Google: [No response, unintelligible phrase partially parsed].

8. User: Empieza una llamada de vídeo.

    Google: ¿A quién quieres hacer una videollamada?

    User: ¿eh?

    Google: [No response, unintelligible phrase partially parsed].

9. User: Crea un listado.

    Google: Vale, ¿Cómo quieres que se llame?

    User: ¿eh?

    Google: [No response, unintelligible phrase partially parsed].

10. User: Manda un correo electrónico.

    Google: No puedo ayudarte a leer ni escribir correos, pero puedo abrir tu aplicación de correo electrónico cuando me lo pidas.